\documentclass[aip,apl,unsortedaddress,superscriptaddress,reprint]{revtex4-1}
\usepackage{hyperref}
\usepackage{graphicx}
\usepackage{amsmath}
\usepackage{amssymb}
\usepackage{times,mathptmx}
\usepackage{dcolumn}
\usepackage{ifthen}
\usepackage{color}
\usepackage{proof}

\newif\ifcom
\newif\ifdel
\comtrue               %
\delfalse

\begin{document}

\title{Current direction anisotropy of the spin Hall magnetoresistance in nickel ferrite thin films with bulk-like magnetic properties} 
\author{Matthias Althammer}
\email{matthias.althammer@wmi.badw.de}
\affiliation{Walther-Mei{\ss}ner-Institut, Bayerische Akademie der Wissenschaften, 85748 Garching, Germany}
\affiliation{Physik-Department, Technische Universit\"at M\"unchen, 85748 Garching, Germany}
\author{Amit Vikam Singh}
\affiliation{University of Alabama, Center for Materials for Information Technology MINT and Department of Chemistry, Tuscaloosa, AL 35487 USA}
\author{Tobias Wimmer}
\affiliation{Walther-Mei{\ss}ner-Institut, Bayerische Akademie der Wissenschaften, 85748 Garching, Germany}
\author{Zbigniew Galazka}
\affiliation{Leibniz-Institut f\"{u}r Kristallz\"{u}chtung, 12489 Berlin, Germany}
\author{Hans Huebl}
\affiliation{Walther-Mei{\ss}ner-Institut, Bayerische Akademie der Wissenschaften, 85748 Garching, Germany}
\affiliation{Physik-Department, Technische Universit\"at M\"unchen, 85748 Garching, Germany}
\affiliation{Nanosystems Initiative Munich (NIM), 80799 M\"unchen, Germany}
\author{Matthias Opel}
\affiliation{Walther-Mei{\ss}ner-Institut, Bayerische Akademie der Wissenschaften, 85748 Garching, Germany}
\author{Rudolf Gross}
\affiliation{Walther-Mei{\ss}ner-Institut, Bayerische Akademie der Wissenschaften, 85748 Garching, Germany}
\affiliation{Physik-Department, Technische Universit\"at M\"unchen, 85748 Garching, Germany}
\affiliation{Nanosystems Initiative Munich (NIM), 80799 M\"unchen, Germany}
\author{Arunava Gupta}
\affiliation{University of Alabama, Center for Materials for Information Technology MINT and Department of Chemistry, Tuscaloosa, AL 35487 USA}

\date{\today}
\begin{abstract}
We utilize spin Hall magnetoresistance (SMR) measurements to experimentally investigate the pure spin current transport and magnetic properties of nickel ferrite (NiFe$_2$O$_4$,NFO)/normal metal (NM) thin film heterostructures. We use (001)-oriented NFO thin  films grown on lattice-matched magnesium gallate substrates by pulsed laser deposition, which significantly improves the magnetic and structural properties of the ferrimagnetic insulator. The NM in our experiments is either Pt or Ta. A comparison of the obtained SMR magnitude for charge currents applied in the [100]- and [110]-direction of NFO yields a change of 50\% for Pt at room temperature. We also investigated the temperature dependence of this current direction anisotropy and find that it is qualitatively different for the conductivity and the SMR magnitude. From our results we conclude that the observed current direction anisotropy may originate from an anisotropy of the spin mixing conductance or of the spin Hall effect in these Pt and Ta layers, and/or additional spin-galvanic contributions from the NFO/NM interface.
\end{abstract}
\maketitle
The advent of (spin) angular momentum transport without an accompanying charge current, i.e.~the flow of pure spin currents, has led to the discovery of several remarkable effects that are relevant for next generation spin electronic devices~\cite{sander_2017_2017,nakata_spin_2017,althammer_pure_2018}. Amongst these effects is the spin Hall magnetoresistance~\cite{Nakayama2013,althammer_quantitative_2013,Hahn2013SMR,Vlietstra2013,chen_theory_2013,Chen2016SMRReview,althammer_pure_2018} in magnetically ordered insulator (MOI)/ normal metal (NM) heterostructures, which enables the detection of novel magnetic phases in MOIs~\cite{aqeel_electrical_2016,Ganzhorn2016,dong_spin_2018}. While initial investigations of the SMR heavily relied on yttrium iron garnet (YIG), the report on the observation of the SMR in many other MOIs ranging from ferrimagnetic~\cite{althammer_quantitative_2013,Isasa2014,hui_spin_2016} to antiferromagnetic~\cite{manchon_spin_2017,hou_tunable_2017,ji_spin_2017,hoogeboom_negative_2017,fischer_spin_2018} order confirms the universality of this effect. The magnitude of the SMR effect crucially depends on the transparency of the MOI/NM interface as well as the spin Hall effect (SHE) and the spin diffusion length of the NM. Nevertheless, the impact of the current direction with respect to the crystalline orientations on the SMR remains up to now unexplored. In this publication we experimentally show that the SMR amplitude in nickel ferrite thin films with bulk-like magnetic properties~\cite{singh_bulk_2017} interfaced with Ta or Pt depends upon the relative orientation of the charge current $\mathbf{j}$ compared to the NFO crystal axes.

The heterostructures investigated in this study are NFO/NM bilayers, where the NM is Pt and Ta. The ferrimagnetic NFO thin films ($\approx 100\;\mathrm{nm}$) are grown on (001)-oriented MgGa$_2$O$_4$ (MgGO) substrates via pulsed laser deposition.
The bulk MgGO single crystals were obtained by the Czochralski method at the Leibniz-Institut f\"{u}r Kristallz\"{u}chtung, Berlin, Germany~\cite{galazka_mgga_2015}, and substrates were then prepared by CrysTec GmbH, Berlin, Germany.
During growth the substrate was kept at $700^\circ\;\mathrm{C}$ in an oxygen atmosphere with $10\;\mathrm{mTorr}$. For the magnetotransport experiments we then defined NM Hall bar structures on top of the NFO with a width of $80\;\mathrm{\mu m}$ and a length of $800\;\mathrm{\mu m}$ via optical lithography, sputter deposition of the NM and lift-off. For the NMs we used Ta and Pt layers, that were deposited ex-situ in an ultra-high vacuum sputtering system with a base pressure of $2\times10^{-9}\;\mathrm{mbar}$. The deposition was carried out in an argon atmosphere at $5\times10^{-3}\;\mathrm{mbar}$ and a growth rate of $2\;\mathrm{{\AA}/s}$ for both materials. The magnetotransport experiments were carried out in two superconducting magnet cryostats at temperatures $T$ ranging from $5\;\mathrm{K}$ to $300\;\mathrm{K}$. One transport setup is based on a 2D-vector magnet with  magnetic fields limited to $\mu_0 H=7\;\mathrm{T}$ and a second has full 3D magnetic field vector control ($\mu_0 H\leq2.5\;\mathrm{T}$). For the resistance measurements we applied a DC current of $10\;\mathrm{\mu A}$ to the Hall bar and measure the longitudinal DC voltage drop. To rule out any spurious thermal voltages, we utilized the current reversal technique as detailed in Refs.~\onlinecite{Schreier2013,Goennenwein2015}.

\begin{figure}[h]
 \includegraphics[width=85mm]{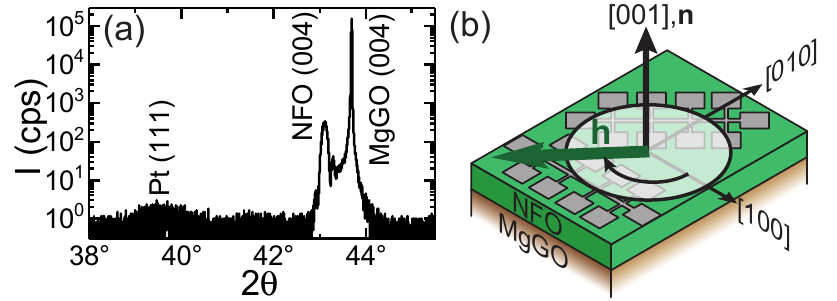}\\
 \caption[XRD results and sample geometry for magnetotransport]{(a) Obtained x-ray diffraction results from a $2\theta$-$\omega$ scan on a $100\;\mathrm{nm}$ thick NFO layer grown on a MgGO substrate and covered with a $10\;\mathrm{nm}$ thick Pt layer. Reflections from the (001)-oriented NFO layer, the MgGO substrate and the (111)-textured Pt are visible. (b) Illustration of the sample geometry used for transport experiments. On top of the NFO layer two NM Hall bar structures are deposited with $\mathbf{j}\parallel\left[100\right]$ and $\mathbf{j}\parallel\left[110\right]$ crystal orientations of the NFO thin film.} 
  \label{figure:XRD_NFO}
\end{figure}
As a first step we investigated the orientation of the sputter deposited Pt layer on top of the NFO layer on (001)-oriented MgGO substrates by x-ray diffraction using a reference sample with a blanket $10\;\mathrm{nm}$ thick Pt film. The obtained results for the $2\theta$-$\omega$ are shown in Fig.~\ref{figure:XRD_NFO}(a). The sample exhibits reflections from the (001)-oriented NFO thin film and the low intensity (111)-reflection of Pt suggests that Pt grows (111)-textured on top of the NFO layer. Due to the low intensity of the Pt reflection, we were unable to investigate any in-plane epitaxial relationship between Pt and NFO. For a Ta reference sample no reflections originating from the Ta layer were found, thus we do not have information on the growth of Ta on our NFO thin films.

\begin{figure*}[t]
 \includegraphics[width=170mm]{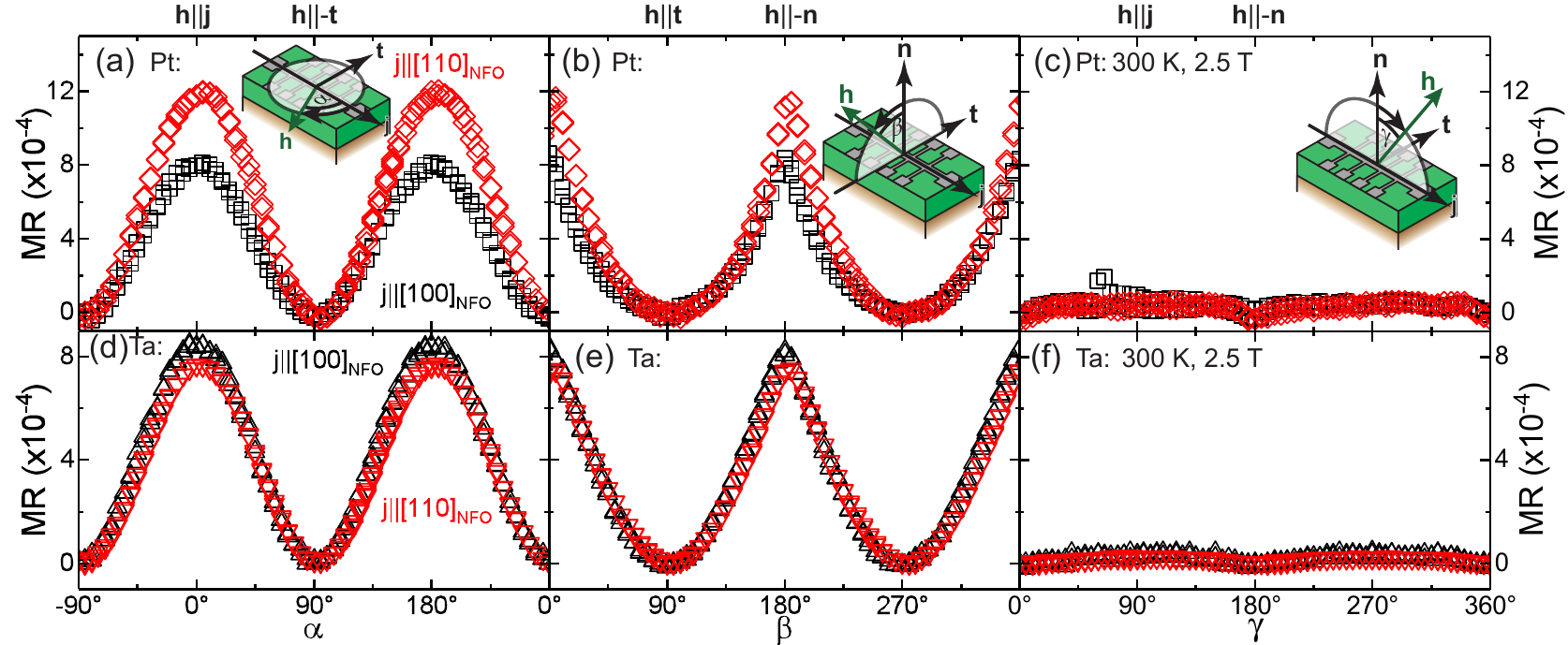}\\
 \caption[ADMR results for Pt and Ta]{ADMR data of a NFO(100)/Pt(3.5) bilayer (a)-(c) and a NFO(100)/Ta(5) bilayer (d)-(f) sample grown on a MgGO (001) substrate. The data has been recorded at $300\;\mathrm{K}$ and an external magnetic field of $2.5\;\mathrm{T}$.  In the plot, black squares and red diamonds represent the experimental data for Pt and black up-triangles and red down-triangles represent the experimental data for Ta for the charge current direction along the [100]- and [110]-direction, respectively. For both materials and both charge current directions, we observe an angle-dependence of the MR in the in-plane and the perpendicular to current direction plane rotations, while negligible angle-dependence is visible in the third orthogonal plane. Thus the observed MR exhibits the symmetry fingerprint of the SMR.}
  \label{figure:ADMR_NFO}
\end{figure*}

For the magnetotransport experiments we utilized two differently oriented NM Hall bars with respect to the crystalline orientation of the NFO layer as illustrated in Fig.~\ref{figure:XRD_NFO}(b). This allows us to investigate the SMR for two different current directions: along the [100]-direction and the [110]-direction of NFO. For the study of the SMR in these samples, we used angle-dependent magnetoresistance (ADMR) experiments~\cite{limmer_angle-dependent_2006}. In ADMR experiments an external magnetic field with fixed magnitude $\mu_0 H$ is applied to the sample, while measuring the longitudinal resistivity $\rho_\mathrm{long}$ of the Hall bar as a function of the orientation of the magnetic field direction $\mathbf{h}=\mathbf{H}/H$. In our experiments we rotated the external magnetic field in several planes to investigate the observed magnetoresistance (MR). The first plane is the in-plane (ip) rotation of the external magnetic field, where $\alpha$ is defined as the angle between the charge current direction and $\mathbf{h}$ (See inset in Fig.\ref{figure:ADMR_NFO}(a)). Four more rotation planes have been used in these experiments. Two rotation planes perpendicular to the two charge current directions (oopj, $\beta$ as defined in the inset of Fig.~\ref{figure:ADMR_NFO}(b)), and two rotation planes residing in the plane defined by each charge current direction and the surface normal (oopt, $\gamma$ as defined in the inset of Fig.~\ref{figure:ADMR_NFO}(c)). 
We determined the minimum value of $\rho_\mathrm{long}$ for each ADMR measurement and calculated the relative MR amplitude as the difference with respect to this minimum value divided by this minimum value.

In Fig.~\ref{figure:ADMR_NFO} we show the ADMR results obtained for Pt and Ta Hall bars on NFO thin films at $300\;\mathrm{K}$ and $\mu_0 H=2.5\;\mathrm{T}$. We first look into the MR response of the Pt Hall bars for the in-plane rotation plane (see Fig.~\ref{figure:ADMR_NFO}(a)). Clearly, for both Hall bars we observe two maxima and two minima over the full $360^\circ$ rotation and the MR follows a $\sin^2$-dependence. For the both current directions we observe maxima in the MR for $\mathbf{h}\perp\mathbf{t}$ and minima for $\mathbf{h}\parallel\mathbf{t}$ in agreement with SMR theory~\cite{chen_theory_2013,Chen2016SMRReview,althammer_pure_2018}. However, the extracted maximum MR for the two current directions is different: For $\mathbf{j}$ along the NFO [100]-direction we find a maximum MR of $8.1\times10^{-4}$, while for $\mathbf{j}$ along the NFO [110]-direction we obtain $1.2\times10^{-3}$. This is a 50\% change in maximum MR for these two current directions and does not originate from a difference in Pt resistivity as discussed below. In addition, the MR for the [110]-direction is comparable to SMR values obtained for YIG/Pt heterostructures, where the current was oriented along the $[1\bar{1}0]$- and the $[1\bar{2}1]$-direction of the YIG film
~\cite{althammer_quantitative_2013,althammer_pure_2018}.

To further investigate wether the increase in maximum MR for the in-plane rotation for $\mathbf{j}$ along the NFO [110]-direction is only due to the SMR, we also conducted ADMR experiments in the oopj- and oopt-configuration for both charge current directions. These results are shown in Fig.~\ref{figure:ADMR_NFO}(b) for the oopj-configuration and in (c) for the oopt-configuration. For the oopj-configuration we see that the MR does not follow a typical $\cos^2$-dependence, which can be explained by the large uniaxial anisotropy with the hard axis along the surface normal for NFO~\cite{singh_bulk_2017}. Thus, for the applied field of $2.5\;\mathrm{T}$ it is not possible to fully align the magnetization direction $\mathbf{m}$ along the out-of-plane direction. Nevertheless, we observe distinct maxima for $\mathbf{h}\perp\mathbf{t}$ and minima for $\mathbf{h}\parallel\mathbf{t}$. Again we find that in the oopj-configuration the maximum MR for $\mathbf{j}$ along the NFO [110]-direction is larger than for $\mathbf{j}$ along the NFO [100]-direction. In the oopt-configuration we only observe a negligible angle-dependence of the MR signal (due to the fact that the magnetic anisotropy in NFO still plays a role at the investigated magnetic field magnitude) for both current directions, in agreement with SMR theory~\cite{chen_theory_2013,althammer_quantitative_2013}. Thus, for both current directions we observe the typical SMR fingerprint in ADMR experiments and can conclude that the observed current direction anisotropy of the MR originates from the SMR. We note that similar results have been obtained in the investigated temperature range from $5\;\mathrm{K}$ to $300\;\mathrm{K}$ for all three rotation planes. Moreover, we conducted the same ADMR experiments on several different NFO/Pt samples and always found this charge current direction anisotropy of the SMR response, such that we can rule out sample thickness variations as well as changes in the resistivity of Pt as the cause for the observed charge current direction dependence.

For comparison, we also investigated the current direction anisotropy in Ta/NFO Hall bar structures. The extracted MR is shown in Fig.~\ref{figure:ADMR_NFO}(d) for the ip-, (e) for the oopj-, and (f) for the oopt-configuration, respectively. Also for Ta we find an angle-dependence of the MR for the ip and oopj-configuration for both current directions and negligible angle-dependence for the oopt-configuration. Thus also for the Ta layer the sole cause for the observed MR is the SMR. In contrast to the Pt Hall bars, the maximum MR is now larger for $\mathbf{j}$ along the NFO [100]-direction ($8.4\times10^{-4}$ for the [100]-direction and $7.7\times10^{-4}$ for the [110]-direction). The difference in the maximum MR for Ta is small and thus also the current direction anisotropy of the SMR.

In order to further investigate this current direction anisotropy of the SMR we conducted ADMR experiments in all three orthogonal rotation planes for temperatures $5\;\mathrm{K}\leq T \leq 300\;\mathrm{K}$ and a maximum external magnetic field $\mu_0 H\leq 7\;\mathrm{T}$. To extract the SMR amplitude from these measurements we simulated the SMR response of $\rho_\mathrm{long}$ using~\cite{chen_theory_2013,Chen2016SMRReview,althammer_pure_2018}:
\begin{equation}
\rho_\mathrm{long}=\rho_0+\rho_1 (1-m_\mathrm{t})^2,\\
\label{Eq:SMRRhoLong}
\end{equation}
where $\rho_0$ is the resistivity of the NM layer, when $\mathbf{m}$ is collinear to the spin polarization of the spin accumulation in the NM layer induced by the SHE. The SMR amplitude is described by $\rho_1$ and $m_\mathrm{t}$ is the projection of $\mathbf{m}$ onto the $\mathbf{t}$-direction ($\mathbf{t}=\mathbf{n}\times\mathbf{j}$). For our simulations, we assumed $\rho_0$ to be field dependent, while $\rho_1$ is field-independent. For the determination of magnetization direction for each field direction we globally optimized the free enthalpy density normalized to the saturation magnetization of the NFO~\cite{limmer_angle-dependent_2006,althammer_quantitative_2013}:
\begin{equation}
G_M(\mathbf{m})=-\mu_0 H(\mathbf{m}\cdot\mathbf{h})+ B_{001}m_\mathrm{001}^2+B_\mathrm{c}(m_\mathrm{100}^4+m_\mathrm{010}^4+m_\mathrm{001}^4),\\
\label{Eq:FreeEnthalpy}
\end{equation}
with $B_{001}$ the uniaxial out-of-plane anisotropy field, $B_\mathrm{c}$ the cubic anisotropy field and $m_{hkl}$ the projection of $\mathbf{m}$ onto the [hkl]-direction of NFO. For each temperature we then optimized a set of $\rho_i$ and $B_i$ parameters until excellent agreement (reduced $\chi^2\leq1\times10^{-6}$) between simulation and experimental data was obtained in all rotation planes and for all $\mu_0 H$. For the cubic magnetic anisotropy of the NFO thin film we found a temperature independent value of $B_\mathrm{c}=10\;\mathrm{mT}$, corresponding to in-plane magnetic easy axes along the [110]-direction and [1$\bar{1}$0]-direction, which agrees with ferromagnetic resonance studies on samples grown under the same conditions~\cite{singh_bulk_2017}.

\begin{figure}[h]
 \includegraphics[width=85mm]{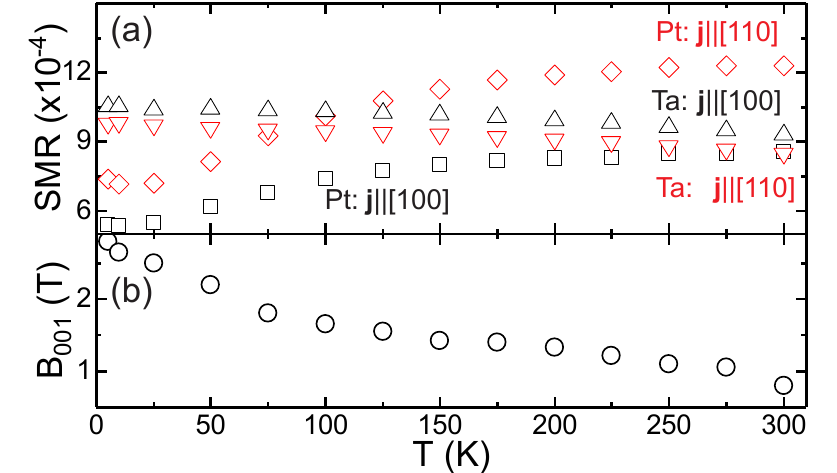}\\
 \caption[Extracted parameter from ADMR experiments]{Extracted simulation parameters from the temperature and field-dependent ADMR experiments. (a) $SMR$ as a function of temperature for Pt with $\mathbf{j}$ along the NFO [100]-direction (black squares), [110]-direction (red diamonds), and for Ta with $\mathbf{j}$ along the NFO [100]-direction (black triangles), [110]-direction (red triangles). Over the whole temperature range the current direction anisotropy of the SMR amplitude persists for Pt and Ta. (b) Evolution of the uniaxial magnetic anisotropy parameter $B_{001}$ with temperature. 
 }
  \label{figure:SMR_Temp}
\end{figure}
To better analyze the temperature dependence of the other parameters we first plot the SMR magnitude $SMR=\rho_1/\rho_0(\mu_0 H=7\;\mathrm{T})$ for Ta and Pt as a function of $T$ for the two different charge current directions  in Fig.~\ref{figure:SMR_Temp}(a). As evident from this plot, the current direction anisotropy persists for all investigated temperatures. For Pt, $SMR$ is larger for $\mathbf{j}$ along the NFO [110] direction over the whole temperature range. At low temperatures ($T\leq 25 \;\mathrm{K}$), the difference in SMR magnitude for the current directions in Pt is smaller than at higher temperatures. For the two Ta Hall bars we find that the $SMR$ is larger for $\mathbf{j}$ along the NFO [100] direction, albeit the difference is less pronounced than for Pt. Moreover, for $T\leq 75 \;\mathrm{K}$ the $SMR$ in Ta is larger than in Pt, suggesting that Ta might be the better choice for SMR investigations at low temperatures.

For the magnetic anisotropy determined from these ADMR experiments, we find that $B_{001}$ monotonically increases with decreasing temperature as illustrated in Fig.~\ref{figure:SMR_Temp}(b). Such a behavior could be either explained by the increase in saturation magnetization or due to strain effects caused by the difference in thermal expansion of the MgGO substrate and the NFO layer. As we do not observe any saturation behavior at low temperatures, which one would expect for the shape anisotropy contribution of a thin film, we conclude that $B_{001}$ is dominated by the strain in the NFO layer 
. This finding agrees well with the previous analysis of the uniaxial out-of-plane magnetic anisotropy in bulk-like NFO thin films~\cite{singh_bulk_2017}.

\begin{figure}[h]
 \includegraphics[width=85mm]{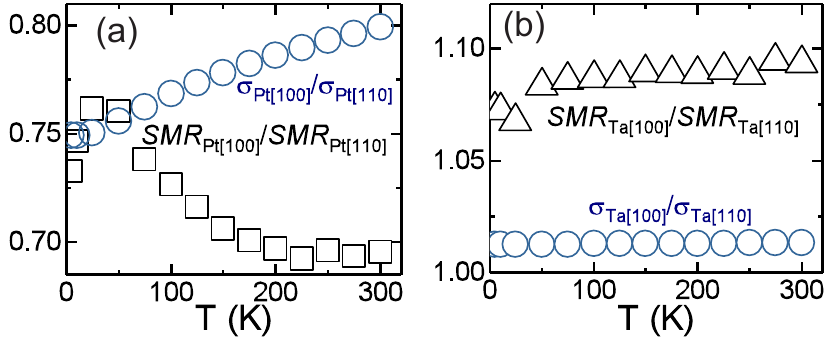}\\
 \caption[Temperature dependence of anisotropy]{Extracted temperature dependence of the ratios of the SMR magnitude $SMR_\mathrm{NM[100]}/SMR_\mathrm{NM[110]}$ (black symbols) and conductivities $\sigma_\mathrm{NM[100]}/\sigma_\mathrm{NM[110]}$ (blue circles) for (a) Pt and (b) Ta. The anisotropy ratios of the SMR and the conductivities exhibit different temperature dependence ruling out any simple correlation between these two anisotropic quantities.}
  \label{figure:SMR_ratio}
\end{figure}
In order to further investigate the origin of the observed current direction anisotropy of the SMR, we compared the temperature evolution of the ratios in $SMR$ to the temperature dependence of the ratio of the conductivities $\sigma$ for the two different current directions for Pt and Ta. The result of this analysis is shown in Fig.~\ref{figure:SMR_ratio}. As evident from Fig.~\ref{figure:SMR_ratio}(a), Pt also exhibits a current direction anisotropy of the conductivity. However, this anisotropy in conductivity cannot be the only reason for the observed $SMR$ current direction anisotropy as the conductivity ratio and the $SMR$ ratio have a quite different dependence on temperature. For $\sigma_\mathrm{Pt[100]}/\sigma_\mathrm{Pt[110]}$ we observe a decrease with decreasing temperature. In contrast, $SMR_\mathrm{Pt[100]}/SMR_\mathrm{Pt[110]}$ shows a more complex non-monotonic temperature dependence. At high temperatures, the $SMR$ ratio remains rather constant at $0.69$, then starts to increase for $T\leq200\;\mathrm{K}$ and reaches a maximum value of $0.76$ for $10\;\mathrm{K}\leq T \leq25\;\mathrm{K}$. From this we conclude that it is not possible to simply correlate the observed current direction anisotropy of the SMR to the 
anisotropy of the conductivity for Pt.

For Ta as illustrated in Fig.~\ref{figure:SMR_ratio}(b), $\sigma_\mathrm{Ta[100]}/\sigma_\mathrm{Ta[110]}$ remains about constant over the whole temperature range with a value of $1.01$. For $SMR_\mathrm{Ta[100]}/SMR_\mathrm{Ta[110]}$, we find a slight decrease with decreasing temperature from $1.09$ at room temperature down to $1.07$ at $T=5\;\mathrm{K}$. Nevertheless, the evolution of these two ratios with temperature is rather different and thus again we can not find an universal relation between the two for Ta.

Even though we carried out temperature- and field-dependent ADMR experiments on the NFO/NM sample, the origin of the current direction anisotropy of the SMR in these samples is difficult to determine. From our experiments we conclude that a texturing of the NM layer seems to increase this anisotropy. In our opinion there are three possible contributions to the SMR that are responsible for the current direction anisotropy. One possibility could be that the spin mixing conductance relevant for the spin current flow across the NFO/NM interface exhibits an anisotropy for the alignment of the spin orientation with respect to the NFO crystal. As the direction of the spin orientation at the NFO/NM interface in the SMR experiments depends on the charge current direction, the transparency of the interface could be different for the two investigated charge current directions and thus the SMR magnitude might also be different. However, to our knowledge such an anisotropy of the spin mixing conductance has never been experimentally observed or theoretically postulated. A second possible mechanism would be an anisotropy of the spin Hall effect in the NM with respect to the charge current direction. Thus, the amount of spin current generated depends on the charge current direction and leads to an anisotropy of the SMR amplitude. Such an anisotropy of the SHE has been theoretically predicted based on ab-initio calculations, but only for materials with hexagonal symmetry~\cite{freimuth_anisotropic_2010}. Last but not least, the bulk spin Hall effect is not the only cause for the conversion of a charge current into a pure spin current in a NM. It is quite possible that contributions from the spin-galvanic effect~\cite{ganichev_spin-galvanic_2002,edelstein_spin_1990,sanchez_spin--charge_2013,jungfleisch_interface-driven_2016,sangiao_control_2015,nakayama_rashba-edelstein_2016,seibold_theory_2017} arising at the NFO/NM interface also give rise to additional contributions to the SMR. As the spin-galvanic effect is caused by spin-orbit fields as a result of the broken inversion symmetry at the NFO/NM interface, it is quite possible that such a contribution depends on the charge current direction. In such a scenario the combined action of bulk SHE and interfacial spin-galvanic effect will determine the SMR magnitude and cause the observed current direction anisotropy, as previously observed in epitaxial Fe/GaAs heterostructures~\cite{chen_robust_2016}. Clearly, determining the origin of the current direction anisotropy requires more sophisticated experiments, which is beyond the scope of this publication. However, our first promising results suggest that understanding this anisotropy in the SMR may provide an additional pathway to engineer the spin current transport across MOI/NM interfaces.

In summary, we showed that the SMR from NFO thin films with bulk-like magnetic properties grown on MgGO substrates interfaced with Ta and Pt is comparable to results obtained on the prototype ferrimagnetic insulator YIG, such that these NFO thin films are well suited for pure spin current experiments in agreement with already published results~\cite{shan_enhanced_2018}. Our results further illustrate that one can change the SMR amplitude and thus also the amount of spin current across the MOI/NM interface by changing the charge current direction. While at the current stage of our investigations we can not pinpoint the physical origin of the observed charge current direction anisotropy, we have to consider at least three possible reasons. First, it may originate from an anisotropy of the SHE in textured or even epitaxial NM layers. Second, spin-galvanic contributions originating from the inversion symmetry breaking at the MOI/NM interface have to be taken into account. Third, an anisotropy of the spin mixing conductance in NFO may explain the observed behavior. From this perspective, further experiments on fully epitaxial MOI/NM systems are expected to allow for a further clarification of possible origins. Moreover, a more systematic investigation of the charge current direction anisotropy in these NFO/Pt bilayers may allow to find a clue to the underlying symmetry of the charge current direction anisotropy of the SMR. Our results presented here open up a new avenue for engineering the charge current to spin current conversion in MOI/NM heterostructures.

We thank Timo Kuschel, Juan Shan and Jutta Schwarzkopf for fruitful discussions. We gratefully acknowledge financial support by the DFG via project AL 2110/2-1. The work at the University of Alabama was supported by NSF Grant No. ECCS-1509875.
\bibliography{Biblio}
\end{document}